\documentclass[a4paper]{article}

\PassOptionsToPackage{hyphens}{url}
\usepackage[bookmarks=false,hidelinks]{hyperref}

\usepackage{INTERSPEECH2020}
\usepackage{epsfig,amssymb,amsmath}
\ninept

\usepackage{bm}

\usepackage[caption=false,font=footnotesize,subrefformat=parens,labelformat=parens]{subfig}
\usepackage{tikz}
\usepackage{pgfplots}
\pgfplotsset{compat=newest}

\def\addlegendimage{\csname pgfplots@addlegendimage\endcsname}

\usepackage{cmap}
\usetikzlibrary{shapes.misc,shapes,mindmap,arrows,arrows.meta,positioning,patterns}
\usepgfplotslibrary{fillbetween}

\usetikzlibrary{patterns.meta}

\tikzset{
    font=\scriptsize
}

\newcommand{\setvariable}[2]{
	\let#1\relax
	\newcommand{#1}{#2}
}

\newcommand*\diff{\mathop{}\!\mathrm{d}}

\newcommand{\given}{\ensuremath\,|\,}

\newcommand{\Cllr}   {\ensuremath{C_{\textrm{llr}}}}

\DeclareMathOperator{\ECE}{ECE}

\renewcommand{\vec}[1]{\bm{#1}}

\usepackage{colorbrewer}
\usepgfplotslibrary{colorbrewer}
\definecolor{darkgreen}{rgb}{0.2,0.7,0.3}
\definecolor{darkorange}{rgb}{1,0.55,0}

\pgfplotsset{
    mated/.style={ybar interval, fill opacity=0.6, area legend, 
        fill=darkgreen, draw=white},
    nonmated/.style={ybar interval, fill opacity=0.6, area legend, 
        fill=white!30!red, draw=white},
    histAxis/.style={
        width=\figwidth,
        height=\figheight,
        scale only axis,
        ymin=0,ymax=1,
        ylabel={Rel. freq.},
        xmajorgrids,ymajorgrids},
    rawHistAxis/.style={histAxis,xlabel={Scores},xmin=-20,xmax=40},
    llrHistAxis/.style={histAxis,xlabel={Oracle LLRs},xmin=-10,xmax=10},
}

\usepackage{eurosym}

\setvariable{\colorScalePurple}{PiYG-11-1}
\setvariable{\colorScaleOrange}{Oranges-9-4}
\setvariable{\colorScaleBlue}{Spectral-5-5}
\setvariable{\colorScaleRed}{hda}
\setvariable{\colorScaleGreen}{Spectral-5-4!80!black}

\usepackage{booktabs}
\usepackage{multirow}
\usepackage{makecell}

\usepackage{balance}

\title{The Privacy ZEBRA: Zero Evidence Biometric Recognition Assessment}

\makeatletter
\def\name#1{\gdef\@name{#1\\}}
\makeatother
\name{{\em Andreas Nautsch$^1$, Jose Patino$^1$, Natalia Tomashenko$^2$, Junichi Yamagishi$^3$,} \\{\em Paul-Gauthier No{\'e}$^2$, Jean-Fran{\c{c}}ois Bonastre$^2$, Massimiliano Todisco$^1$ and Nicholas Evans$^1$}}

\address{$^1$ Digital Security Department, EURECOM, France \\
$^2$ Laboratoire Informatique d'Avignon (LIA), Avignon Universit{\'e}, France\\
$^3$ National Institute of Informatics, 2-1-2 Hitotsubashi, Chiyoda-ku, Tokyo, Japan}

\email{nautsch@eurecom.fr}
\begin{document}
\maketitle

\begin{abstract}
Mounting privacy legislation calls for the preservation of privacy in speech technology, though solutions are gravely lacking.
While evaluation campaigns are long-proven tools to drive progress, the need to consider a privacy \emph{adversary} implies that traditional approaches to evaluation must be adapted to the assessment of privacy and privacy preservation solutions. 
This paper presents the first step in this direction:
\emph{metrics}.

We introduce the zero evidence biometric recognition assessment \mbox{(ZEBRA)} framework and propose 
two new privacy metrics.
They measure the \emph{average} level of privacy preservation afforded by a given safeguard for a population and the 
\emph{worst-case} privacy disclosure
for an individual.  
The paper demonstrates their application to privacy preservation assessment within the scope of the VoicePrivacy challenge.
While the ZEBRA framework is designed with speech applications in mind, it is a candidate for incorporation into biometric information protection standards and is readily extendable to the study of privacy in applications even beyond speech and biometrics.

\end{abstract}

\section{Introduction}

Spoken language contains a wealth of personal information including the biometric identity~\cite{nautsch2019gdpr,Kroeger2020}.  
Such sensitive information is clearly susceptible to being exploited for unscrupulous and ethically reprehensible purposes.  Unsurprisingly,  speech data falls within the scope of recent privacy legislation, e.g., the  California Consumer Privacy Act (CCPA) \cite{CaliforniaState-CCPA-2018} in the US (effective January 1, 2020), the European General Data Protection Regulation (GDPR) \cite{EU-GDPR-2016} (implemented May 25, 2018) and EU directive 2016/680 \cite{EU-PoliceDirective-2016} (the Police Directive, also implemented May 25, 2018).  Privacy is a fundamental human right \cite{EU-HumanRights-2010} and the failure to respect privacy legislation can attract significant fines. 
Speech technology providers and operators are thus obliged to ensure adequate provisions for privacy preservation. 

There are two general approaches to guard against privacy intrusions in the case of speech data. The first is to protect access to speech data, usually via some form of encryption and secure computation. The second, and the focus in this article, is to strip the speech signal of personally identifiable information such that it cannot be linked (with some level of certainty) to a specific individual. Pseudonymisation, de-identification and anonymisation are all examples of such approaches, but are all relatively embryonic research topics within the speech community; only few solutions have been proposed thus far.

One reason for why progress in privacy preservation has not kept pace with legislation is due to the lack of frameworks for assessment.  While privacy preservation solutions may take many different forms, their goals are common: they should prevent the use of speech being used to infer identity.  Accordingly, the same or similar metrics can be used for the assessment of any form of privacy preservation solution.  The design of such metrics is then a priority and stands to boost progress in privacy preservation whatever the particular approach. 

Upon first consideration, metrics for the assessment of privacy preservation solutions may seem straightforward.  This is not the case, however, since many obvious metrics 
do not reflect the \emph{decision policy of a privacy adversary}.  Consequently, they will give a misleading measure of privacy.  Inspired by metrics used in forensics research, this paper reports our proposals for two different privacy preservation metrics that disentangle the considerations of the privacy safeguard and the privacy adversary.  The paper shows how the \emph{empirical cross entropy} and the \emph{strength of evidence} can be harnessed within a so-called \emph{zero evidence biometric recognition assessment} (ZEBRA) framework to measure the expected and worst-case privacy disclosure for a given privacy preservation solution. 

The motivation for this work is described in Section~\ref{sec::motivation}.  Section~\ref{sec:ece} describes background work and the empirical cross-entropy. Section~\ref{sec:zebra} describes its use in the ZEBRA framework and demonstrates its application to the assessment of privacy preservation solutions within the context of the VoicePrivacy 2020 challenge~\cite{Tomashenko-VoicePrivacy-EvalPlan-2020,Tomashenko-VoicePrivacy-Interspeech-2020}.  A discussion of the work and directions for the future are presented in Section~\ref{sec:conclusion}.

\section{Towards empirical privacy metrics}
\label{sec::motivation}

Privacy metrics are needed in order to gauge and to compare the level of privacy preservation offered by different solutions.  Such a metric should also reflect the \emph{gain in privacy} delivered by a given safeguard as well as the \emph{remaining potential for privacy disclosure}.  
We also seek metrics which reflect not just the \emph{average} level of privacy preservation provided by a particular solution, but one that can also be used to understand the \emph{variation} in privacy preservation provided to a population; there may be differences in the level of protection provided to different users.  Finally, metrics should be based not upon the prior beliefs and costs of a privacy preservation system designer or evaluator but should, instead, reflect those of a privacy adversary. Only then, can we gain meaningful insights to privacy and the gap to \emph{perfect privacy}.

\emph{Perfect privacy} was introduced as \emph{perfect secrecy} by Shannon~\cite{Shannon-SecrecyTheory-1949}: the posterior probabilities of intercepted data are identical to the prior probabilities of an adversary. This led to \emph{theoretically} founded assessment of privacy safeguards in modern cryptography~\cite{lindell2014introduction} (\emph{zero knowledge proofs}); here, input data has a \emph{mathematical definition}. Speech data is different (we use \emph{models}, not \emph{definitions}), hence we seek \emph{empirical} approaches to assessment. Unfortunately, despite some obvious candidates, existing empirical metrics do not meet the above requirements.

\begin{figure}[t]
    \centering
    \begin{tikzpicture}
        \node[draw=black] (pp) {Safeguard};
        \node[draw=black, right=4em of pp] (c) {Classifier};
        \node[draw=black, right=3em of c] (d) {Decision policy};
        \node[draw=black, right=2em of d] (r) {Action};
        \draw[-latex] (pp.east) -- node[above,midway] (data) {Protected} node[below,midway] {data} (c.west);
        \draw[-latex] (c.east) -- node[above,midway] (scores) {Scores} (d.west);
        \draw[-latex] (d) -- node[above,midway] (action) {} (r);
        \node[below=.25em of c] (error) {Error trade-offs};
        \node[below=.25em of d] {Inference information};
        \node[below=.25em of r] (impact) {Impact};
        \draw [decorate,decoration={brace,amplitude=4pt}] ([yshift=.76em]pp.north west|-scores.north) -- node [black,above,midway,yshift=0.5em]  {Realm of \emph{privacy preservation}} ([yshift=0.75em]scores.north);
        \draw [decorate,decoration={brace,amplitude=4pt}] (data|-scores.north) -- node [black,above,midway,yshift=.5em,xshift=2em]  {Realm of the \emph{adversary}} (impact.east|-scores.north);
    \end{tikzpicture}
    \caption{Decoupling the classifier, decision policy and action to estimate privacy from the perspective of an adversary.}
    \vspace{-0.2cm}
    \label{fig:decoupling_layers}
\end{figure}
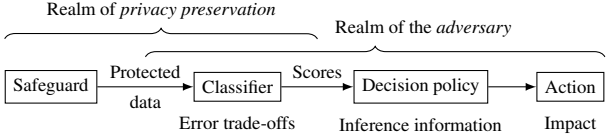

Obvious candidates are to measure the, e.g., equal error rate (EER), \emph{unlinkability} of protected biometric datasets \cite{marta18unlinkMetric}, cost risks and application-independent risks by detection cost functions (DCFs)~\cite{Doddington-NISTSpeakerRecognitionEvaluation-SD-2000} and the goodness of so-called log-likelihood ratio (LLR) scores, $\Cllr$~\cite{BrummerDuPreez-ApplicationIndependentEvaluation-SpeakerDetection-ComputerSpeechLanguage-2006}\footnote{
    The $\Cllr$ metric also relates to empirical cross-entropy (see Section~\ref{sec:ece}) for one specific prior; \emph{a computational co-incidence} \cite{RamosGonzalezRodriguez-CrossEntropyInformation-ForensicSpeakerRecognition-Odyssey-2008}.
}; all are ill suited to the privacy scenario.\footnote{
    (1) 
    By its very definition, 
    the EER reflects a privacy adversary’s worst possible decision policy~\cite{Brummer-deVilliers-BOSARIS-Binary-Scores-AGNITIO-Research-2011}; the EER will hence reflect an unduly optimistic estimate of privacy protections. (2) Unlinkability considers the potential of evidence to \emph{confirm} an identity, it overlooks the relevance to privacy of evidence that might \emph{exclude} an identity. The exclusion of identities is still evidence, however, and so the unlinkability metric also gives a potentially distorted measure of privacy protections. (3) For cost based metrics, the impact of privacy disclosure depends on subcultures, such that we cannot possibly treat cost impacts.
}
The envisaged scenario is illustrated in Fig.~\ref{fig:decoupling_layers}. Speech data is processed according to some form of privacy preservation algorithm (safeguard) to suppress speaker-discriminative information.  The resulting data is used by an adversary who still seeks to infer the speaker's identity.
The adversary does this using some biometric classifier which assesses data and reports on the \emph{strength-of-evidence} in the form of scores.  In this scenario, the classifier can be in the realm of either the privacy preserver, or the privacy adversary. 
If the privacy adversary were to use the best technology available to them, then a classifier can be used within the realm of the privacy preserver so long as it is representative of the state of the art.  This is where the realm of the privacy preserver ends.

From here on, everything is within the realm of the privacy adversary, not the privacy preserver.  A privacy metric must hence reflect the adversary's decision policy 
(what action to take) 
parameterised by their prior and cost beliefs.  It is then necessary to assume minimum knowledge, or maximum uncertainty.  
Whereas priors can be simulated, costs of privacy infringement cannot.  In this case, the \emph{prior of the prior} is assumed to be uniformly distributed.
Accordingly, classifier outputs are assumed to be likelihood ratios, obtained by the \emph{pool adjacent violators to LLR algorithm}~\cite{Brummer-deVilliers-BOSARIS-Binary-Scores-AGNITIO-Research-2011,brummer09PAVDemonstration,Brummer-MeasuringRefiningCalibrating-SpeakerLanguageInformation-UniStellenbosch-2010}, a non-linear transform of observed scores resulting in \emph{oracle score calibration}.

The scenario described above may be familiar to some readers for it resembles that of a (counter) forensics scenario.   The counterparts here are the forensic practitioner (the privacy preserver) and the judge/jury (the privacy adversary).  The forensic practitioner must present evidence only; s/he must not make decisions.  The decision policy is that of the judge/jury; it is unknown to the forensic practitioner who must assume maximum uncertainty of prior/cost beliefs so as not to encroach upon the province of the court. While they provide the basis for the work presented in this paper, even the metrics used by the forensics community are not sufficient for the assessment of privacy.  This is because the forensic practitioner simulates the empirical cross-entropy over different prior values only, without quantifying either the expected or highest strength-of-evidence (worst-case disclosure), which we need for assessing privacy.

\section{Empirical cross entropy}
\label{sec:ece}

This section presents a brief overview of empirical cross entropy (ECE), a metric developed for the assessment of forensic speaker recognition systems and originally reported in~\cite{RamosGonzalezRodriguez-CrossEntropyInformation-ForensicSpeakerRecognition-Odyssey-2008,Ramos-ForensicEvaluation-UPM-2007,ramos18crossEntropy}.  It ends with a discussion of how the ECE can be adapted to assess the performance of privacy preserving systems.

A speaker recognition system is furnished with two utterances, one for training with known speaker identity and one for testing an identity claim.
There are two propositions $\mathcal{A}, \mathcal{B}$, namely that either the identity of the speakers in each utterance is the same $\mathcal{A}$, or that the speakers are different $\mathcal{B}$. We denote the set of propositions $\Theta= \{\mathcal{A}, \mathcal{B}\}$ and refer to a single proposition as $\theta \in \Theta$. Before using the speaker recognition system, we have some prior belief in the truth of each proposition, which we quantify by the prior probability $P(\theta)$. We denote two probability spaces: 
the ground truth or \emph{reference} probability $P(\cdot)$, 
and the \emph{classifier} probability $\tilde{P}(\cdot)$; the latter is the forecast whose desired value is the reference. The prior entropy  
in making a binary decision $H_P(\Theta)$ is:

\begin{equation}
    H_P(\Theta) = -\sum_{\theta\in\Theta} P(\theta)\log_2 P(\theta).
\end{equation}

The recognition system compares the training utterance to the test utterance in order to compute (ideally calibrated) recognition scores $S$. 
The goal of an attacker is to reduce \emph{prior posterior entropy} about the propositions $\Theta$ by updating the prior with observed scores $S$ which results in the \emph{posterior entropy}, see \cite{Cover-Thomas-Elements-Information-Theory-Wiley-2006}. 
The reference \emph{posterior entropy} 
$H_P(\Theta \given S)$ is:

\begin{equation}
    H_P(\Theta \given S) = -\sum_{\theta\in\Theta} P(\theta) \int_s P(s \given \theta)\log_2 P(\theta\given s)\diff{s}.
\end{equation}

Because we are in an empirical setting, it is not possible to derive reference likelihoods $P(s \given \theta)$ from a theoretical footing; in general, they remain unknown  \cite{RamosGonzalezRodriguez-CrossEntropyInformation-ForensicSpeakerRecognition-Odyssey-2008,ramos18crossEntropy}. 
We can, however, quantify the \emph{cross-entropy} $H_{P||\tilde{P}}(\Theta \given S)$ between the posterior distributions of a system $\tilde{P}$ and the reference $P$:
\begin{equation}
    \hspace{-.125em}{H}_{P||\tilde{P}}(\Theta \given S) = -\sum\limits_{\theta \in \Theta} P(\theta)\int_s P(s \given \theta)\log_2 \tilde{P}(\theta \given s) \diff{s}.
\end{equation}

\noindent We can approximate $P(s \given \theta)\approx |S_{\theta}|^{-1}$ for a \emph{large number} of classifier posteriors $\tilde{P}(\theta \given s)$, where $S_{\theta}$ denotes the set of scores for class $\theta$ and $|S_{\theta}|$ is its size. In the forensic setting, systems do not compute $\tilde{P}(\theta \given s)$ directly, since priors $\tilde{P}(\theta)$ are decoupled.  The choice of reference and classifier prior values $P(\theta), \tilde{P}(\theta)$ is external to the classifier and considered a parameter: $\pi = P(\mathcal{A}) = \tilde{P}(\mathcal{A})$ and $1-\pi  = P(\mathcal{B}) = \tilde{P}(\mathcal{B})$. Consequently, systems estimate likelihood ratio (LR) scores $S=\frac{\tilde{P}(X \given \mathcal{A})}{\tilde{P}(X \given \mathcal{B})}$ from features $X$.
The ECE is computed as \cite{RamosGonzalezRodriguez-CrossEntropyInformation-ForensicSpeakerRecognition-Odyssey-2008}:
\begin{equation}
    \begin{aligned}
        \hspace{-.5em}\ECE(\Theta \given \mathcal{S}) &:= \frac{\pi}{|\mathcal{S}_{\mathcal{A}}|} \sum_{a \in \mathcal{S}_{\mathcal{A}}}\log_2\left(1 + \frac{1-\pi}{a\,\pi}\right) \\
        &+\frac{1-\pi}{|\mathcal{S}_{\mathcal{B}}|}\sum_{b \in \mathcal{S}_{\mathcal{B}}}\log_2\left(1 + \frac{b\,\pi}{1-\pi}\right).
    \end{aligned}
    \label{eq:ECEplot}
\end{equation}

It can be shown that the ECE reflects the expected amount of additional information that is needed in order to know the true proposition $\theta$.
If the classifier is unreliable (it performs poorly, requiring more information), then the ECE will be higher than if the classifier is more reliable.

\section{Zero evidence framework}
\label{sec:zebra}

The goal of privacy preservation is to strip a speech utterance of personally identifiable information such that an adversary cannot recognise the identity of a speaker from an protected recording of their voice. 
In terms of speaker recognition, it should not be possible to match with certainty a training utterance to an anonymised test utterance.  
The adversary has some prior belief and seeks to use evidence provided by a speaker recognition system to update their prior belief for identity inference.   
Since use of a speaker recognition system should not result in an information gain, privacy preservation should leave the adversary in a position where they
are left making decisions based only upon their prior belief (whatever it is). 
This is {\it perfect secrecy}~\cite{Shannon-SecrecyTheory-1949}, a concept re-coined here as {\it perfect privacy}.

This section sets out our ideas to make use of the ECE as a means of assessing the level of privacy provided by a privacy preserving solution.  We propose two metrics that can be used for optimisation, assessment and ranking according to a so-called \emph{zero evidence biometric recognition assessment} approach.\footnote{
    Code: \url{https://gitlab.eurecom.fr/nautsch/zebra}
}  
The two metrics aim to measure the extent to which a safeguard preserves privacy, or rather what degree of speaker discriminative information remains in an utterance.  
The first metric reflects the gain in information that an adversary can obtain by using a speaker recognition system. This is equivalent to the expected privacy disclosure regardless of the adversary's prior belief.  Realising that the ECE reflects no more than an expected value (\emph{population} level), the second metric reflects the worst-case scenario (\emph{individual} level), i.e.\ the maximum level of privacy that may be disclosed despite privacy preservation. 

\subsection{Expected privacy disclosure}\label{sec:proposed:ECE}

\begin{figure}[t]
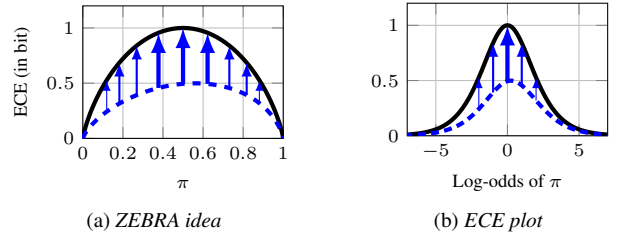

    \centering
    \vspace{-1em}
    \setvariable{\figwidth}{75pt}
    \setvariable{\figheight}{50pt}
    \subfloat[ZEBRA idea]{\input{figures/ece0_idea}}
    \hfill
    \subfloat[ECE plot]{\input{figures/ece_plot}}
    \caption{(a) Adversary ECE (blue profile) for some oracle calibrated scores. The idea is to make the blue equal the black profile (perfect privacy ECE). (b) Conventionally, ECE profiles are plotted against $\log\frac{\pi}{1-\pi}$, the log-odds of $\pi$.}
    \label{fig:sre-plots:eceIdea}
\end{figure}

The expected privacy disclosure is 
the relative information that can be gained from use of a speaker recognition system.  This difference is illustrated in terms of the ECE profiles in Fig.~\ref{fig:sre-plots:eceIdea} where the black line represents the \emph{perfect privacy} ECE (\emph{zero evidence} scores $\bm{0}_{\mathcal{S}}$; all LRs have the value $1$) 
and the blue, dashed line represents the \emph{adversary} ECE, i.e.\ for a system that would yield oracle calibrated scores $\mathcal{S}$ (oracle LRs).\footnote{In \cite{RamosGonzalezRodriguez-CrossEntropyInformation-ForensicSpeakerRecognition-Odyssey-2008,Ramos-ForensicEvaluation-UPM-2007,ramos18crossEntropy},  the \emph{default} and the \emph{minimum} ECE, respectively.}

By removing as much as possible any 
biometric information in
speech data, the 
gap between these two profiles would reduce and result in an increase in the adversary ECE, as indicated by the blue arrows in Fig.~\ref{fig:sre-plots:eceIdea}.  In the case that the safeguard is successful in removing \emph{all} the speaker specific information, then the perfect privacy and adversary ECEs would be identical, i.e.\ we have \emph{perfect secrecy}; there remains \emph{zero evidence}.\footnote{ $\mathrm{ECE}(\Theta\given \bm{0}_{\mathcal{S}}) = \pi\log_2(1+\frac{1-\pi}{\pi}) + (1-\pi)\log_2(1+\frac{\pi}{1-\pi})$.}

In practice, a safeguard is unlikely to remove \emph{all} evidence; 
it will likely still result in the disclosure of some privacy and different solutions will disclose different levels of privacy.  Hence, we need some means to compare solutions.  The answer is a metric which measures the difference between the \emph{perfect privacy} and \emph{adversary} ECE.
Since the adversary's prior is unknown, the evaluator must assume maximum uncertainty (different $\pi$ values have the same probability of occurrence).
Accordingly, the metric must reflect the difference between both ECE profiles 
in Fig.~\ref{fig:sre-plots:eceIdea} for the full range of priors $\pi$.
This gives the \emph{expected privacy disclosure} $D_{\textrm{ECE}}(\Theta \given \mathcal{S})$:
\begin{equation}
    \begin{aligned}
        D_{\textrm{ECE}}(\Theta \given \mathcal{S}) &=
        \int_0^1\ECE(\Theta \given \bm{0}_{\mathcal{S}}) - \ECE(\Theta \given \mathcal{S})\diff{\pi}.
    \end{aligned}
    \label{equ:D_ECE}
\end{equation}

It can be shown that the integral is given by:

\begin{equation}
    \begin{aligned}
        D_{\textrm{ECE}}(\Theta \given \mathcal{S}) &= \frac{\langle Z(a)\rangle_{a\in\mathcal{S}_{\mathcal{A}}} + \langle Z(\frac{1}{b})\rangle_{b\in\mathcal{S}_{\mathcal{B}}}}{\log(2)}\\
        \text{with}\quad Z(x) &= \frac{(x-3)(x-1)+2\log(x)}{4\,(x-1)^2},\\
        x>0,\qquad &\lim\limits_{x\mapsto 1}Z(x) = 0,
        \quad \text{and}\quad \lim\limits_{x\mapsto\infty}Z(x) = \frac{1}{4},
    \end{aligned}
    \label{eq::d_ece}
\end{equation}

\noindent where $\langle\cdot\rangle_{s \in \mathcal{S}_{\theta}}$ computes the average over scores $s$ in a set $\mathcal{S}_{\theta}$.

$D_{\textrm{ECE}}(\Theta \given \mathcal{S})$ signifies the expected disclosure to an adversary in bits and
is independent of the adversary's prior belief (the prior is marginalised and integrated out in Eq.~\ref{equ:D_ECE}).  It has an intuitive interpretation: if $D_{\textrm{ECE}}(\Theta \given \mathcal{S}) = \frac{1}{2\,\log(2)} \approx 0.721$, then classes are perfectly separated for all priors (\emph{no privacy}); if $D_{\textrm{ECE}}(\Theta \given \mathcal{S}) = 0$, then we have \emph{perfect privacy} (\emph{zero evidence}).

\subsection{Worst-case privacy disclosure}

The worst-case privacy disclosure is reflected by the highest strength of evidence observed during testing.
While we can use posterior probabilities of a non-informative prior $\pi$ as in~\cite{Nautsch-PrivacySpeaker-SpeechCommunication-181203}, we propose the use of an additional approach which improves human interpretability, especially for the non-expert. 
This entails \emph{categorical tags}, see Tab.~\ref{tab:ece-tags}, for LR values $l$ that are adapted from methodological guidelines for forensic practitioners published by the European Network of Forensic Science Institutes~\cite{Drygajlo-MethodologicalGuidelinesForensics-ENFSI-2015} and by forensics work in~\cite{Nordgaard-Scale-of-conclusion-Law-Probability-Risk-2012}. Additionally, we provide the odds ratio for recognising a biometric identity by the lowest LR value of a category (assuming no knowledge of the adversary prior: $\pi$ is flat); 
lower odds mean less precision for an adversary, thus privacy is more preserved (best is $50:50$).

\begin{table}[b]
    \centering
    \caption{Categorical tags of worst-case privacy disclosure.}\vspace{-.25em}
    \label{tab:ece-tags}
    \setlength{\tabcolsep}{5pt}
    \begin{tabular}{ccc}
        \toprule
        Tag & Category & Posterior odds ratio (flat prior)\\
        \midrule
        0 & $\hphantom{10}l = 1 = 10^0$ & $50:50$ (flat posterior)\\
        \midrule
        A & $10^0 < l < 10^1$ & more disclosure than $50:50$\\
        B & $10^1 \leq l < 10^2$ & one wrong in 10 to 100\\
        C & $10^2 \leq l < 10^4$ & \hphantom{0\,}one wrong in 100 to 10\,000 \\
        D & $10^4 \leq l < 10^5$ & one wrong in 10\,000 to 100\,000\\
        E & $10^5 \leq l < 10^6$ & \,one wrong in 100\,000 to 1\,000\,000\\
        F & $10^6 \leq l \hphantom{{{}<{}10^6}}$ & \hphantom{000}one wrong in at least 1\,000\,000\\
        \bottomrule
    \end{tabular}
\end{table}

To determine $l$, the LR value of the worst-case privacy disclosure, we must consider both positive strength of evidence (proposition $\mathcal{A}$, it \emph{is} the speaker) and negative strength of evidence (proposition $\mathcal{B}$, it is \emph{not} the speaker).  
The former corresponds to LR values $\mathcal{A}: 1<l<\infty$, whereas the latter corresponds to LR values $\mathcal{B}: 0<l<1$.  
In LR space, however, the extent to which one proposition is favoured over another is non-linear; LRs of $l=0.9$ or $l=1.1$ do not reflect the same strength of evidence in support of each proposition.
A linear space is obtained easily by operating in the log-LR (or LLR) space $\mathcal{S}'_{\log}$ of ideally calibrated\footnote{
    We refer to $\mathcal{S}'$ instead of $\mathcal{S}$ on purpose. To avoid infinite LR values, see the code of \cite{Brummer-deVilliers-BOSARIS-Binary-Scores-AGNITIO-Research-2011}, we extend the calibration used in Section~\ref{sec:proposed:ECE}, and apply \emph{Laplace's rule of succession} (also known as \emph{the sunrise problem}). Two dummy scores are added to the extremities of all observed scores---one for class $\mathcal{B}$ and one for class $\mathcal{A}$. The former serves as a Bayesian predictor to the LR value for the highest class $\mathcal{A}$ scores (that are larger than the highest class $\mathcal{B}$ score), and the latter captures the infinity. 
} scores $\mathcal{S'}$:
\begin{equation}
    \mathcal{S}'_{\log} = \left(\log(s) \given s \in \mathcal{S'}\right).
\end{equation}

In log-LR 
space, a value of \emph{zero} implies neither $\mathcal{A}$ nor $\mathcal{B}$ is favoured (zero evidence; full privacy).

In addition, scores in log-LR 
space are symmetric; scores of -0.1 and +0.1 reflect the same strength of evidence for each proposition.  
As a result, the worst-case privacy disclosure is obtained as $\log(l)$ by taking the maximum of the absolute value:\footnote{
    The metric $\log(l)$ corresponds to the $L^{\infty}$ length-norm of a vector $\bm{x}$ with $n$ LLRs: $||\bm{x}||_{\infty}=\max\{|x_1|, \dots, |x_n|\}$. If this length is zero, there is \emph{zero evidence} disclosure (the vector points from its origin to itself). A privacy disclosure, a non-zero dimension, increases the vector length. The worst-case metric is \emph{optimistic}, since any other $L^p$ norm with $p<\infty$ is larger (despite the Bayesian prediction of LLRs).
}
\begin{align}
    \log(l) &= \max_{s\in\mathcal{S}'_{\log}}(\mathrm{abs}(s)).
    \label{eq::log_l}
\end{align}

As for log-LRs, 
the log-odds of posteriors/priors are also
symmetric; Tab.~\ref{tab:ece-tags} defines categories by magnitudes---for base~10 log-LRs as $\log_{10}(l)$ values and posterior log-odds.

\subsection{Assessing privacy disclosure, an example}
\label{sec:voiceprivacy}

Reported here is an example case study performed using the \mbox{ZEBRA} framework in the context of the \emph{VoicePrivacy 2020} challenge~\cite{Tomashenko-VoicePrivacy-EvalPlan-2020,Tomashenko-VoicePrivacy-Interspeech-2020}, which involves the design of anonymisation solutions in privacy preservation. Experiments were performed with the female subset of the LibriSpeech test data using: the two challenge baselines, B1 \cite{fang2019speaker} (a pre-trained x-vector approach) and B2 \cite{Patino-McAdams-EURECOM-2019} (a formant shifting technique; no training)---the \emph{safeguard} component in Fig.~\ref{fig:decoupling_layers}; a state-of-the-art x-vector~\cite{snyder2018x} speaker recognition system---the \emph{classifier} component in Fig.~\ref{fig:decoupling_layers}.

\begin{figure}[t]
    \centering
    \vspace{.5em}
    \setvariable{\figwidth}{75pt}
    \setvariable{\figheight}{40pt}
    \input{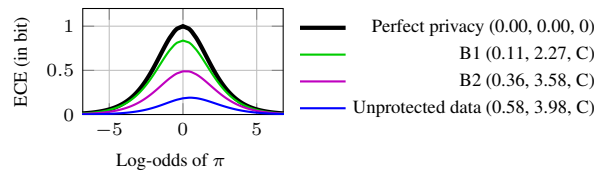}
    \caption{ZEBRA assessment with ECE profiles.}
    \label{fig:ece:vp2020}
\end{figure}

Results are illustrated in Fig.~\ref{fig:ece:vp2020}.  
The legend also shows ZEBRA results in the form of a $(D_{\textrm{ECE}}$, $\log_{10}(l),$ tag) tuple, where computations are according to \eqref{eq::d_ece} for $D_{\textrm{ECE}}$ and to \eqref{eq::log_l} for $\log_{10}(l)$, with categorical tags referenced in Tab.~\ref{tab:ece-tags}.  The lower blue profile shows the ECE without protection ($D_{\textrm{ECE}}$: 0.58, $\log_{10}(l)$: 3.98, tag: C).  The magenta and green profiles show the ECE of each baseline. The black profile corresponds to perfect privacy or \emph{zero evidence}.  

The baselines offer varying degrees of privacy preservation. B1 ($D_{\textrm{ECE}}$: 0.11) appears to perform considerably better than B2 ($D_{\textrm{ECE}}$: 0.36); the green and black profiles are relatively close together whereas the gap between black and magenta profiles is substantial.  
$\log_{10}(l)$ results show a somewhat different picture: 3.98 for the unprotected system, 3.58 for B2 and 2.27 for B1.  These results \emph{all} correspond to tag C in Tab.~\ref{tab:ece-tags}.\footnote{
    Categorical tags for LRs and their tables evolved over time since 1961 starting with Jeffrey, whose table ended with $2 \leq \log_{10}(l)$ \cite{Kaye-Weight-of-Evidence-NIST-TC-Forensics-2016}; category C resulted after the introduction of DNA evidence in media.
}  Hence, while B1 is substantially better than B2 in terms of \emph{average} privacy disclosure, there is less to choose between them in terms of the categorical \emph{worst-case} privacy disclosure.

These observations highlight the value of the proposed \mbox{ZEBRA} framework for privacy preservation. Whereas the EER and other metrics may give an unduly optimistic and distorted sense of protection, the $D_{\textrm{ECE}}$ provides a  reliable estimate of protection in terms of the \emph{average} protection afforded to a \emph{population} whereas $\log(l)$ provides additional insights into the level of protection afforded in a \emph{ worst-case} to an \emph{individual}.

\section{Discussion and future work}
\label{sec:conclusion}

With mounting legislation demanding protections for personal information, provisions for privacy are today of paramount consideration.  This paper presents the ZEBRA framework for the evaluation of privacy safeguards. It overcomes many of the weaknesses of existing metrics such as the equal error rate. Crucially, the framework formulates the problem in the realm of a privacy adversary and provides the means to assess the average and worst-case privacy disclosure of a given safeguard.

While the work 
makes inroads towards an adversary-centric metric, the arguments presented in this paper suggest that we need to go one step further; metrics are only one part of an assessment strategy.  The case study presented in this paper uses a protocol designed by an \emph{evaluator}.  Just like the decision policy, the protocol is also in the realm of the \emph{adversary}. 
Future work should hence extend the current study to consider 
the privacy that can be disclosed when the adversary chooses both the decision policy \emph{and} the protocol.
Similarly, the current work assumes oracle score calibration,
whereas it too is in the realm of the adversary. 
Future work could hence study the impact of calibration within the ZEBRA framework, such that they cannot be used by adversaries.

Finally, the ZEBRA framework proposed in this paper is relevant to the \emph{biometric information protection} standard, currently in revision.  
It is also readily extendable to the study of privacy concerning sensitive speech data, e.g.\ health and emotional status, or particularly sensitive spoken/transcribed content, e.g.\ political and religious beliefs.
Finally, speech serves as only one example application of the ZEBRA framework; since it operates in the score domain, it can be applied with minimal effort to the study of non-speech problems such as privacy preservation in video surveillance.

\section{Acknowledgements}

This work is partly funded by the projects: ANR-JST VoicePersonae, ANR Harpocrates and ANR-DFG RESPECT.

\newpage
\balance
\bibliographystyle{IEEEtran}
\bibliography{main}

\begin{thebibliography}{10}
\providecommand{\url}[1]{#1}
\csname url@samestyle\endcsname
\providecommand{\newblock}{\relax}
\providecommand{\bibinfo}[2]{#2}
\providecommand{\BIBentrySTDinterwordspacing}{\spaceskip=0pt\relax}
\providecommand{\BIBentryALTinterwordstretchfactor}{4}
\providecommand{\BIBentryALTinterwordspacing}{\spaceskip=\fontdimen2\font plus
\BIBentryALTinterwordstretchfactor\fontdimen3\font minus
  \fontdimen4\font\relax}
\providecommand{\BIBforeignlanguage}[2]{{%
\expandafter\ifx\csname l@#1\endcsname\relax
\typeout{** WARNING: IEEEtran.bst: No hyphenation pattern has been}%
\typeout{** loaded for the language `#1'. Using the pattern for}%
\typeout{** the default language instead.}%
\else
\language=\csname l@#1\endcsname
\fi
#2}}
\providecommand{\BIBdecl}{\relax}
\BIBdecl

\bibitem{nautsch2019gdpr}
A.~Nautsch, C.~Jasserand, E.~Kindt, M.~Todisco, I.~Trancoso, and N.~Evans,
  ``The {GDPR} \& speech data: Reflections of legal and technology communities,
  first steps towards a common understanding,'' in \emph{Proc. Interspeech},
  2019, pp. 3695--3699.

\bibitem{Kroeger2020}
J.~L. Kr{\"o}ger, O.~H.-M. Lutz, and P.~Raschke, ``Privacy implications of
  voice and speech analysis -- information disclosure by inference,'' in
  \emph{Proc. Privacy and Identity Management. Data for Better Living: AI and
  Privacy: IFIP Int'l Summer School 2019, Revised Selected Papers}.\hskip 1em
  plus 0.5em minus 0.4em\relax Springer, 2020, pp. 242--258.

\bibitem{CaliforniaState-CCPA-2018}
{California State Legislature}, ``Assembly bill no. 375, chau. privacy:
  personal information: businesses ({C}alifornia {C}onsumer {P}rivacy {A}ct),''
  6 2018.

\bibitem{EU-GDPR-2016}
{European Council}, ``Regulation 2016/679 of the {E}uropean {P}arliament and of
  the {C}ouncil on the protection of individuals with regard to the processing
  of personal data and on the free movement of such data ({G}eneral {D}ata
  {P}rotection {R}egulation),'' 4 2016.

\bibitem{EU-PoliceDirective-2016}
------, ``Directive 2016/680 of the {E}uropean {P}arliament and of the
  {C}ouncil on the protection of individuals with regard to the processing of
  personal data by competent authorities for the purposes of the prevention,
  investigation, detection or prosecution of criminal offences or the execution
  of criminal penalties, and on the free movement of such data, and repealing
  {C}ouncil {F}ramework {D}ecision {2008/977/JHA},'' 4 2016.

\bibitem{EU-HumanRights-2010}
{Council of Europe}, ``European convention on human rights,'' 6 2010.

\bibitem{Tomashenko-VoicePrivacy-EvalPlan-2020}
N.~Tomashenko, B.~M.~L. Srivastava, X.~Wang, E.~Vincent, A.~Nautsch,
  J.~Yamagishi, N.~Evans, J.~M. Patino, J.-F. Bonastre, P.-G. No{\'e}, and
  M.~Todisco, ``The {VoicePrivacy} 2020 challengeevaluation plan,'' 2020.

\bibitem{Tomashenko-VoicePrivacy-Interspeech-2020}
------, ``Introducing the {VoicePrivacy} initiative,'' in \emph{Proc.
  Interspeech}, submitted.

\bibitem{Shannon-SecrecyTheory-1949}
C.~E. Shannon, ``Communication theory of secrecy systems,'' \emph{Bell System
  Technical Journal}, vol.~28, no.~4, pp. 656--715, 10 1949.

\bibitem{lindell2014introduction}
J.~Katz and Y.~Lindell, \emph{Introduction to modern cryptography}.\hskip 1em
  plus 0.5em minus 0.4em\relax Chapman and Hall/CRC, 2014.

\bibitem{marta18unlinkMetric}
M.~Gomez-Barrero, J.~Galbally, C.~Rathgeb, and C.~Busch, ``General framework to
  evaluate unlinkability in biometric template protection systems,'' \emph{IEEE
  Trans. on Information Forensics and Security (TIFS)}, vol.~3, no.~6, pp.
  1406--1420, Jun. 2018.

\bibitem{Doddington-NISTSpeakerRecognitionEvaluation-SD-2000}
G.~R. Doddington, M.~A. Przybocki, A.~F. Martin, and D.~A. Reynolds, ``The
  {NIST} speaker recognition evaluation - overview, methodology, systems,
  results, perspective,'' \emph{Elsevier Science Speech Communication},
  vol.~31, pp. 225--254, 6 2000.

\bibitem{BrummerDuPreez-ApplicationIndependentEvaluation-SpeakerDetection-ComputerSpeechLanguage-2006}
N.~Br{\"u}mmer and J.~du~Preez, ``Application-independent evaluation of speaker
  detection,'' \emph{Elsevier Computer Speech and Language (CSL)}, vol.~20,
  no.~2, pp. 230--275, 7 2006.

\bibitem{RamosGonzalezRodriguez-CrossEntropyInformation-ForensicSpeakerRecognition-Odyssey-2008}
D.~Ramos and J.~Gonzalez-Rodrigues, ``Cross-entropy analysis of the information
  in forensic speaker recognition,'' in \emph{Proc. IEEE Odyssey}, 2008.

\bibitem{Brummer-deVilliers-BOSARIS-Binary-Scores-AGNITIO-Research-2011}
N.~Br{\"u}mmer and E.~{de Villiers}, ``The {BOSARIS} toolkit user guide:
  Theory, algorithms and code for binary classifier score processing,''
  [Online] \url{https://sites.google.com/site/bosaristoolkit}, accessed
  2020-01-15, AGNITIO Research, South Africa, Tech. Rep., 12 2011.

\bibitem{brummer09PAVDemonstration}
N.~Br{\"u}mmer and J.~du~Preez, ``The {PAV} algorithm optimizes binary proper
  scoring rules,'' 2009.

\bibitem{Brummer-MeasuringRefiningCalibrating-SpeakerLanguageInformation-UniStellenbosch-2010}
N.~Br{\"u}mmer, ``Measuring, refining and calibrating speaker and language
  information extracted from speech,'' Ph.D. dissertation, University of
  Stellenbosch, 2010.

\bibitem{Ramos-ForensicEvaluation-UPM-2007}
D.~Ramos-Castro, ``Forensic evaluation of the evidence using automatic speaker
  recognition systems,'' Ph.D. dissertation, Universidad Poli{\'e}cnica de
  Madrid, 2007.

\bibitem{ramos18crossEntropy}
D.~Ramos, J.~Franco-Pedroso, A.~Lozano-Diez, and J.~Gonzalez-Rodriguez,
  ``Deconstructing cross-entropy for probabilistic binary classifiers,''
  \emph{Entropy}, vol.~20, no.~3, p. 208, 3 2018.

\bibitem{Cover-Thomas-Elements-Information-Theory-Wiley-2006}
T.~M. Cover and J.~A. Thomas, \emph{Elements of Information Theory},
  2nd~ed.\hskip 1em plus 0.5em minus 0.4em\relax John Wiley \& Sons, 2006.

\bibitem{Nautsch-PrivacySpeaker-SpeechCommunication-181203}
A.~Nautsch, A.~Jimenez, A.~Treiber, J.~Kolberg, C.~Jasserand, E.~Kindt,
  H.~Delgado, M.~Todisco, M.~A. Hmani, A.~Mtibaa, M.~A. Abdelraheem, A.~Abad,
  F.~Teixeira, D.~Matrouf, M.~Gomez-Barrero, D.~Petrovska-Delcr{\'e}taz,
  G.~Chollet, N.~Evans, T.~Schneider, J.-F. Bonastre, B.~Raj, I.~Trancoso, and
  C.~Busch, ``Preserving privacy in speaker and speech characterisation,''
  \emph{Computer Speech and Language, Special issue on Speaker and language
  characterization and recognition: voice modeling, conversion, synthesis and
  ethical aspects}, vol.~58, pp. 441--480, 11 2019.

\bibitem{Drygajlo-MethodologicalGuidelinesForensics-ENFSI-2015}
A.~Drygajlo, M.~Jessen, S.~Gfroerer, I.~Wagner, J.~Vermeulen, and T.~Niemi,
  ``Methodological guidelines for best practice in forensic semiautomatic and
  automatic speaker recognition including guidance on the conduct of
  proficiency testing and collaborative exercises,'' Europen Network of
  Forensic Science Institutes, Tech. Rep., 2015.

\bibitem{Nordgaard-Scale-of-conclusion-Law-Probability-Risk-2012}
A.~Nordgaard, R.~Ansell, W.~Drotz, and L.~Jaeger, ``Scale of conclusions for
  the value of evidence,'' \emph{Law, Probability and Risk}, vol.~11, no.~1,
  pp. 1--24, 2012.

\bibitem{fang2019speaker}
F.~Fang, X.~Wang, J.~Yamagishi, I.~Echizen, M.~Todisco, N.~Evans, and J.-F.
  Bonastre, ``Speaker anonymization using {X-vector} and neural waveform
  models,'' in \emph{Proc. Speech Synthesis Workshop (SSW)}, 2019, pp.
  155--160.

\bibitem{Patino-McAdams-EURECOM-2019}
\BIBentryALTinterwordspacing
J.~{P}atino, M.~{T}odisco, A.~{N}autsch, and N.~{E}vans, ``{S}peaker
  anonymisation using the {M}c{A}dams coefficient,'' Eurecom, Tech. Rep.
  Research Report RR-20-343, 02 2020. [Online]. Available:
  \url{http://www.eurecom.fr/publication/6190}
\BIBentrySTDinterwordspacing

\bibitem{snyder2018x}
D.~Snyder, D.~Garcia-Romero, G.~Sell, D.~Povey, and S.~Khudanpur, ``X-vectors:
  Robust {DNN} embeddings for speaker recognition,'' in \emph{2018 IEEE
  International Conference on Acoustics, Speech and Signal Processing
  (ICASSP)}, 2018, pp. 5329--5333.

\bibitem{Kaye-Weight-of-Evidence-NIST-TC-Forensics-2016}
D.~H. Kaye, ``The weight of evidence in law, statistics, and forensic
  science,'' in \emph{Proc. NIST Technical Colloquium on Quantifying the Weight
  of Forensic Evidence}, 2016, [Online]
  \url{https://www.nist.gov/system/files/documents/2020/01/22/03_kaye_16-nist-woe-linear.pdf},
  2020-05-06.

\end{thebibliography}

\end{document}